\documentclass[screen]{acmart}

\AtBeginDocument{%
  }

\setcopyright{acmlicensed}

\usepackage[utf8]{inputenc}
\usepackage{array}
\usepackage{multirow}
\usepackage{booktabs}

\usepackage{graphicx} 

\usepackage{ragged2e} 

\usepackage{geometry} 
\usepackage{pgfplots}
\pgfplotsset{compat=1.17} 
\usepackage{caption}
\usepackage{subcaption}
\usepackage{tikz}

\setcopyright{acmlicensed}
\copyrightyear{2025}
\acmYear{2025}
\acmDOI{XXXXXXX.XXXXXXX}
\acmConference[Conference '25]{The Fifth ACM Conference}{November 5–7, 2025}{PA}
\acmISBN{978-1-4503-XXXX-X/2018/06}

\begin{document}

\title{Digital Labor: Challenges, Ethical Insights, and Implications}

\author{ATM Mizanur Rahman}
\affiliation{
  \institution{University of Illinois Urbana-Champaign}
  \city{Urbana}
  \state{Illinois}
  \country{USA}}
\email{amr12@illinois.edu}

\author{Sharifa Sultana}
\affiliation{
  \institution{University of Illinois Urbana-Champaign}
  \city{Urbana}
  \state{Illinois}
  \country{USA}}
\email{sharifas@illinois.edu}

\renewcommand{\shortauthors}{Rahman et al.}

\begin{abstract}
Digital workers on crowdsourcing platforms (e.g., Amazon Mechanical Turk, Appen, Clickworker, Prolific) play a crucial role in training and improving AI systems, yet they often face low pay, unfair conditions, and a lack of recognition for their contributions. To map these issues in the existing literature of computer science, AI, and related scholarship, we selected over 300 research papers on digital labor published between 2015 and 2024, narrowing them down to 143 on digital gig-labor for a detailed analysis. This analysis provides a broad overview of the key challenges, concerns, and trends in the field. Our synthesis reveals how the persistent patterns of representation and voices of gig workers in digital labor are structured and governed. We offer new insights for researchers, platform designers, and policymakers, helping them better understand the experiences of digital workers and pointing to key areas where interventions and future investigations are promptly needed. By mapping the findings from the past ten years' growth of the domain and possible implications, this paper contributes to a more coherent and critical understanding of digital labor in contemporary and future AI ecosystems.

\end{abstract}



\received{20 February 2007}
\received[revised]{12 March 2009}
\received[accepted]{5 June 2009}

\maketitle

\section{Introduction}

The rapid growth of artificial intelligence (AI) and online platforms has created new ways for tedious and laborious tasks to be completed. Many companies now use crowdsourcing platforms like Amazon Mechanical Turk to complete small yet voluminous tasks that they do not want to invest in with their human and machine labor in \cite{irani2015cultural}. These platforms depend on digital workers who perform tasks such as labeling data or reviewing content. These tasks are important for improving AI systems and sustaining data-driven markets; however, this crucial component --- the invisible source of labor, e.g., the workers --- often face various challenges, including low pay, unfair working conditions, and zero or inadequate recognition for their contributions \cite{1-silberman2018responsible, 2-irani2015difference, 9-tandon2022barriers}. Over the past few years, researchers have started to study the lives and challenges of these digital workers. They have looked at concerns like how algorithms control workers, the lack of fairness and transparency in the digital labor market, and how particularly digital gig workers are often treated as invisible or replaceable \cite{2-irani2015difference, 12-kaun2020shadows, 22-png2022tensions}. Some studies have also focused on how digital gig workers organize themselves to demand better treatment or how platform governance can be improved to support more equitable conditions. Despite this growing interest, there are still many systemic inefficiencies and structural inequalities that have not been fully studied or solved. For example, there remains limited cross-cutting synthesis of how platform dynamics, regulatory frameworks, and worker strategies interact across varied global contexts. Additionally, the domain is yet to fully understand how these challenges affect workers in different countries or from different socioeconomic backgrounds, impacting equitable representation within the crowdsourced industry. This leaves a significant gap in our understanding and makes it harder to create fair, accountable, and sustainable solutions that address market inefficiencies and inequities.

We contribute to understanding digital labor and laborers, the challenges digital gig workers face, and the different ethical issues associated with them. To support this goal, we examined a wide range of research papers from different areas, including computer science, HCI, and AI ethics, which helped us bring together scattered insights and build a more complete picture of the digital labor landscape. In total, we reviewed 143 research papers published between 2015 and 2024, focusing on digital gig workers and their experiences. This review provides a broad overview of the key challenges, systemic issues associated with machines, algorithms, humans, and policies, and ethical concerns within the field of digital labor. Additionally, we found recurring patterns, unresolved challenges, and underexplored areas within the digital labor literature. This review also surfaced that certain topics, such as power asymmetries, data invisibility, and platform accountability, were frequent and core to many different studies. However, the lessons from these studies are yet to be addressed in practice in the digital labor market.

This paper makes four key contributions to the human-centered AI and digital labor literature: 
\begin{quote}
\textbf{First,} we map the growth of digital gig labor scholarship in the past ten years and present the key challenges, systemic issues, and sociotechnical concerns associated with the invisible workforce that runs and maintains AI and other data-driven platforms. \\
\textbf{Second,} we highlight how task allocations, algorithmic decision-making, and labor protections are shaped by the interplay of Western-dominated technological infrastructures and worker-led adaptations, making non-Western gig laborers vulnerable to the conflicting and ineffective geopolitical policies that barely function in their favor. \\
\textbf{Third,} through the mapping, we surface the significant gaps and concerns in the scholarship that have been poorly and barely addressed in research and in practice in professional settings. \\
\textbf{Fourth,} our discussions and proposed implications offer insights for researchers, platform designers, and policymakers in coordinating fair, accountable, and more sustainable digital labor ecosystems for gig workers and other stakeholders.
\end{quote}

Note that not all tasks on crowd platforms are associated with AI, at least at the time when this manuscript was written. However, many major AI systems use crowds for different tasks within their data-driven processes and systems. Therefore, this paper brings in the formally outsourced work to crowd, intended tasks performed by users of different services for incentives and rewards, and unintended tasks imposed on humans (e.g., forced ad-views) under the umbrella term ``digital labor", and concerns about the digital gig labor that is associated with different AI systems and infrastructures.


\section{Related Work}

Literature review papers are important because they provide a comprehensive understanding of a field's growth and identify key trends, challenges, and opportunities. For example, one study reviewed human-computer interaction for development papers published between 2009 and 2014 and showed how research expanded to new regions and topics while also identifying collaboration challenges between researchers and practitioners \cite{dell2016ins}. Another review analyzed migration and displacement in human-centered computing research articles published between 2010 and 2019, highlighting expanding the domain's focus from addressing urgent needs toward exploring deeper political and emotional aspects of migration \cite{sabie2022decade}. Similarly, a systematic review of social computing research studying social media (2008–2020) categorized key phases and common methods and revealed the gaps like limited use of feminist approaches \cite{shibuya2022mapping}. These and other similar reviews of domain-specific scholarship are valuable because they summarize large bodies of work, make it easier to understand how a field evolves and adapts to global challenges, and indicate how their growth needs to be navigated in a futuristic yet sustainable way. They also highlight gaps in knowledge and inspire future research directions crucial for advancing and introducing responsible practices in the field. 

Inspired by the above-mentioned literature review research, we attempt similar mapping for studies at the intersection of digital labor, crowdsourced markets, and ethical concerns around data and algorithmic systems. While some previous reviews explored narrower aspects of crowdsourcing—such as its ethical, managerial, or technological dimensions—this paper offers a broader review of research focusing on the conditions, market structures, and governance mechanisms shaping digital labor. Recent research has also emphasized the role of stakeholder engagement and participatory practices in designing more inclusive and accountable AI and platform systems, reflecting growing efforts to address ethical and structural gaps \cite{eaamo-kallina2024stakeholder, eaamo-delgado2023participatory, eaamo-sloane2022participation, eaamo-birhane2022power}. We identify patterns in how digital labor is governed and experienced, and we draw on this body of work to better understand the ethical, economic, and structural gaps that persist in the field. In this section, we discuss two key areas. First, we outline prior research on labor and politics in the gig economy. Then, we examine how data and algorithmic practices are shaping the future of digital labor, including challenges related to fairness, accountability, market inefficiencies, and long-term sustainability.

\subsection{Labor and Politics in Gig Economy}

Gig work refers to short-term, flexible jobs often facilitated through online platforms, where workers are not regular employees and lack traditional employment benefits \cite{gig-spreitzer2017alternative, gig-watson2021looking, 66-wu2024gig}. Surrounding this gig work is an economic system, known as the gig economy, where platforms connect workers with clients for short-term jobs, offering flexibility but often lacking job security \cite{gige-donovan2016does, gige-koutsimpogiorgos2020conceptualizing, gige-roy2020future, 58-stewart2017regulating}. While the gig economy offers flexible work opportunities, but these often come with poor working conditions and a lack of legal protections. Gig workers face significant challenges such as low wages, scams, and limited rights. Studies have shown that platforms frequently control workers through algorithmic management, creating a system where workers must follow strict rules rather than gain empowerment  \cite{lit1-hickson2024freedom, lit3-gerber2021community}. Platforms also contribute to dishonesty, including deceptive pricing and identity falsification, which highlights power imbalances in platform labor \cite{lit5-grohmann2022platform}. Digital platforms prioritize efficiency over fairness, further exacerbating inequality and uncertainty for workers \cite{lit4-kuhn2021human}. While these systems claim to promote freedom, they often mask significant challenges and domination faced by gig workers \cite{lit1-hickson2024freedom, lit2-hansson2014micro}. More recent studies have explored how participation and feedback mechanisms could reshape power relations between workers and platforms, emphasizing the need for genuine, context-specific, and non-extractive forms of worker involvement \cite{eaamo-sloane2022participation, eaamo-birhane2022power, eaamo-corbett2023power, eaamo-russo2024bridging}.

Prior literature review studies on crowdwork have explored specific aspects of crowdsourcing, including its ethical, managerial, and technological dimensions. These reviews provide valuable insights into the evolving field of crowdsourcing, focusing on diverse themes such as ethics, learning opportunities, and process frameworks. Durward \cite{136-durward2016there} conducted a literature review on the ethical dimensions of crowdsourcing using the PAPA framework (privacy, accuracy, property, and accessibility). This review highlighted the lack of attention to workers’ individual perspectives, particularly their working conditions and benefits. Bhatti \cite{137-bhatti2020general} presented a survey of 234 articles and introduced a three-step framework for crowdsourcing processes, including task design, task implementation, and answer aggregation, emphasizing the need for improved task management, evaluation, and worker incentives. Drechsler \cite{138-drechsler2025systematic} systematically reviewed the learning opportunities available within the crowdworkers' workplace, identifying factors like digital tools, platform interfaces, and community interactions as key determinants of learning experiences. Bazaluk \cite{139-bazaluk2024crowdsourcing} examined the intersection of crowdsourcing and migration, performing a thematic analysis to understand how migrant workers engage with crowdsourcing platforms. The study concluded that crowdsourcing could effectively connect migrant workers with employers but recommended a critical analysis of its implications. Other recent works have also emphasized the unique struggles faced by rural and marginalized communities in crowd work, suggesting more tailored platform and market designs to support different populations \cite{hcomp-flores2020challenges, 41-abbas2022goal, hcomp-imteyaz2024human}. We build upon these reviews and extend the discourse by focusing on digital workers and the systems that govern their labor. We highlight the ethical, economic, and structural challenges workers face across diverse contexts, paying particular attention to their treatment, rights, and equitable representation within platform-based markets.

\subsection{Data, Algorithm, and Future of Labor in Crowd}

To support the sustainable growth of platform-based labor, it is essential that it ensures fairness and accountability of its data-driven processes and algorithmic infrastructure, as well as supports the workers in the long run. These workers form the foundation of many AI-driven and data-intensive platforms, but the systems that support their long-term well-being and growth are still weak or fragmented \cite{eaamo-okolo2024explainable}. For digital labor to remain viable, the surrounding infrastructure must help workers not just keep their jobs, but also grow with changing demands and opportunities. That’s why it is essential to examine how problems like unfair pay, biased task distribution, and the absence of platform responsibility are intertwined, possibly making it harder for workers to perform to their fullest potential. Without recognizing how these problems are connected and their deeper causes and effects, attempts to improve platform market dynamics may still remain insufficient. This subsection highlights the need for such mapping as a foundation for designing stronger, more future-ready gig-labor infrastructure that can support the growth and stability of the digital gig-labor industry while reducing systemic inefficiencies.

The literature consistently highlights unfair treatment in digital labor, particularly in how workers are paid, evaluated, and included in platform governance. Workers from economically vulnerable regions are often paid below minimum wage standards, and their contributions go unrecognized \cite{10-irani2023algorithms, 109-haralabopoulos2019paid, 55-salminen2023fair}. Platforms use task-based payment structures that create income instability and exclude hidden labor such as time spent learning tools or waiting for work \cite{111-berg2015income, 112-chen2024we}. Studies also show that algorithms assign tasks in biased ways, giving fewer opportunities to workers based on their region or profile \cite{117-zhang2017consensus, 119-eickhoff2018cognitive}. Moreover, many workers report having little control over their workflow, with limited communication or input into how their work is structured or assessed \cite{13-grohmann2021beyond, 116-ford2015crowdsourcing}. These issues show that fairness in pay, access, and treatment is still missing in many current platform markets. This lack of fairness creates real harm for workers. It leads to constant stress about income, difficulty planning for the future, and a sense of being excluded from decision-making. When people do not feel respected or supported, it becomes harder to stay motivated or continue working on these platforms long-term. Building on these concerns, newer discussions about data labor and synthetic data practices have added further worries about how workers’ contributions are appropriated, often without clear consent or benefit sharing, revealing regulatory gaps in data-driven industries \cite{faact-li2023dimensions, faact-whitney2024real}. These concerns highlight the ongoing need to rethink labor fairness at a structural and policy design level.

Platform accountability is another major concern raised in the literature. Workers can be suspended without notice, underpaid without recourse, or ignored when reporting grievances \cite{120-grossman2018crowdsourcing, 121-zou2018proof, 16-mcinnis2016running}. Despite these harms, platforms rarely face consequences because there are no independent oversight systems in place \cite{13-grohmann2021beyond}. While guidelines have been proposed to improve platform responsibility, such as routine audits, transparency reports, or third-party reviews, few of these systems have actually been put into practice \cite{16-mcinnis2016running, 17-mcinnis2017crowdsourcing, eaamo-barker2023feedbacklogs}. This lack of structural accountability creates fear and vulnerability among workers, who often depend on platforms as a primary income source. The literature shows that without institutional mechanisms to hold platforms responsible, worker concerns are left unresolved.
This creates a power imbalance where platforms can act without consequence, while workers have no way to respond or appeal. It discourages workers from speaking up, even when they are treated unfairly. Over time, this can lead to distrust in the system and a lack of confidence that things will improve. Without real accountability, even well-meaning reforms may fail to protect those who need them most.

Long-term worker well-being and career development are rarely supported by existing crowd platforms. Most digital labor consists of short-term tasks with limited learning opportunities or career mobility \cite{17-mcinnis2017crowdsourcing, 25-naude2022crowdsourcing}. Even though many workers want to grow their skills, access to tutorials, certifications, or advancement tools is limited \cite{109-haralabopoulos2019paid}. As a result, workers face unstable income, emotional fatigue, and burnout, and these challenges make digital labor unsustainable over time \cite{13-grohmann2021beyond, 46-de2022understanding}. Researchers have called for platform models that promote worker development and long-term participation, yet these recommendations are often absent in real-world design \cite{111-berg2015income, 55-salminen2023fair}. This creates a cycle where workers remain stuck in low-paying roles with no clear path forward. Many are forced to treat this work as temporary, even when they rely on it for survival. The lack of growth opportunities also limits the overall quality of the workforce, as experienced workers leave and new ones struggle to adapt. Without investment in skill-building, equitable representation, and career support mechanisms, the entire crowdsourced market becomes fragile, harming both workers and platforms themselves.

Although existing research has explored issues of fairness, sustainability, and accountability in crowd-based digital labor, these areas are often treated in isolation. The broader scholarship frequently discusses problems like unfair pay or algorithmic bias, but often fails to explain how these fairness-related challenges lead to deeper sustainability risks, such as emotional burnout or worker dropout. Similarly, while some studies address sustainability or fairness, they often overlook how the absence of platform accountability allows these problems to persist. The literature rarely maps how these concerns are interconnected with each other. Our work addresses this gap by systematically tracing how challenges in fairness cascade into sustainability breakdowns, and how these are reinforced by weak or missing accountability structures. By surfacing these overlooked connections, our review offers a broader and more integrated understanding of how platform labor systems function. This understanding is essential for designing systems where platform workers can grow, adapt, and sustain their roles over time within a supportive and evolving infrastructure \cite{faact-widder2024epistemic}.

\section{The 'Worker' Terminologies}

\begin{table}[t!]
\centering
\begin{tabular}{|p{2.2cm}|p{12.8cm}|}
\hline
\multicolumn{1}{|c|}{\textbf{Term}} & \multicolumn{1}{c|}{\textbf{Definition}} \\ \hline
\textbf{Crowdworker} & 
A person who performs tasks via an online platform that connects them to requesters or clients. Crowdworkers typically engage in short-term, task-based work, often without a direct employment relationship with the platform or client (definition borrowed from International Labour Organization (ILO) and Kittur et al. \cite{151-CrowdworkGigEconomy, 152-kittur2013future}). \\ \hline
\textbf{Ghost worker} & 
An individual whose labor is essential to the functioning of digital systems but remains invisible to end-users. Ghost workers perform tasks such as data cleaning, content moderation, and AI training under conditions that often lack visibility, fair compensation, or recognition (definition borrowed from Gray et al., and Berg et al. \cite{gray2019ghost, 34-berg2018digital}). \\ \hline
\textbf{Crowdsourcing platforms} & 
Online platforms that act as intermediaries, connecting clients with workers to perform tasks ranging from data annotation to graphic design. Examples include Amazon Mechanical Turk, Upwork, and Clickworker. These platforms often automate task assignment and evaluation while minimizing direct interaction between workers and clients (definition borrowed from Ipeirotis et al., and Deng et al. \cite{155-ipeirotis2010analyzing, 127-deng2016duality}). \\ \hline
\textbf{Freelancer} & 
A self-employed individual offering services, often through online marketplaces or directly to clients, typically with greater autonomy over task selection and work arrangements than crowdworkers (definition borrowed from Friedman et al. \cite{157-friedman2014workers}). \\ \hline
\end{tabular}
\caption{Definitions of the key terminologies related to digital labor.}
\label{tab:definitions}
\end{table}

There are major distinctions between the different terminologies that describe various types of workers in digital labor. Based on existing literature and industry definitions, we explain terms such as crowd worker, ghost worker, freelancer, and crowdsourcing platforms. To ensure consistency, we adopt definitions from widely recognized sources such as the International Labour Organization (ILO) and academic studies on digital labor and summarize them in Table~\ref{tab:definitions}.

This classification simplifies a range of working conditions and experiences, which we clarify using definitions from recognized sources. For instance, a crowdworker on one platform may have significant autonomy and higher pay, while another may face precarious conditions similar to those of ghost workers. The rights and recognition of these workers often depend on their classification. Freelancers and some crowdworkers may negotiate terms and have legal protections in certain jurisdictions, while ghost workers frequently lack these rights due to their invisible status. Similarly, workers on crowdsourcing platforms often operate under strict terms dictated by platform algorithms, with limited recourse for grievances. For the rest of the article, we use the term “crowdworker” to refer broadly to individuals performing tasks on crowdsourcing platforms, unless specific distinctions are necessary. The term “ghost worker” is used when emphasizing the hidden nature of labor. The word “platforms” refers to the intermediaries facilitating this labor. When discussing both categories collectively, we use the term “digital workers.” Through these terminologies, we aim to highlight the diverse experiences and challenges faced by workers in the digital economy.

\section{Mapping the Literature}



This section first describes the methods and strategies we used to review research papers. We wanted to show the trajectory of the domain over the past ten years, so for our review we chose papers published between 2015 and 2024. While it would’ve been nice to have an overview of the domain’s timeline from the beginning to date, it is impossible to put a marker on when exactly the domain officially took off. So, we focused on the last ten years because this period reflects the exponential growth of crowd-based digital labor, especially in the context of cross-border economies and platform-driven work. We initially selected over 300 research papers on digital labor and then narrowed them down to 143 for a detailed analysis. As we continued collecting papers, we noticed that new results were becoming less relevant. We found that most of the important papers surfaced within the first 300, and beyond that, there were still some relevant ones, but they appeared much less frequently—perhaps one in every thirty papers. This filtering helped us focus on studies that provided the most meaningful insights into the field. Then, we summarize the key findings that highlight important insights into the realities of digital work, including the challenges and opportunities.

\subsection{Method and Strategy}





\begin{figure}[t!]
  \centering
  \includegraphics[scale=0.6]{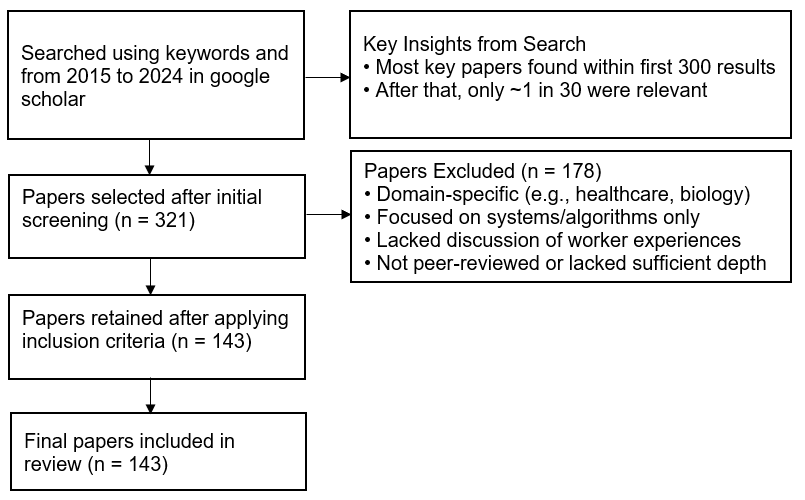}
  \caption{Diagram illustrating the paper selection process for this literature review.}
  \label{4-flowchart}
\end{figure}

We began our literature survey by searching for papers on Google Scholar to identify research articles, reports, and other relevant publications focusing on digital labor and workers from 2015 to 2024. Our search employed a range of terms, including “Crowdwork,” “Ghost work,” “Crowdsourcing,” “Crowdsource,” “Crowdplatform,” and “Ghost Workers,” ensuring comprehensive coverage of the topic. The search fields included titles, abstracts, keywords, and full-text bodies of the documents. Table ~\ref{search} summarizes our literature survey's search parameters. This initial search yielded over 300 publications. We then manually reviewed the documents to assess their relevance to our research focus. During this process, we applied a set of inclusion and exclusion criteria to ensure the quality and relevance of the selected literature. We excluded papers that concentrated only on domain-specific applications of crowdsourcing, such as in healthcare, medicine or bioinformatics, as well as those that focused primarily on system design, algorithms, or theoretical models without discussing worker experiences or labor conditions. We retained only peer-reviewed papers that provided sufficient depth and insight into digital labor. Based on this filtering process, we selected 143 papers for further analysis. Figure~\ref{4-flowchart} summarizes the full selection process from initial retrieval to final inclusion.

We developed and used a rubric of key information to organize and analyze the literature systematically, capturing details such as "when", "where", "who", "what", "why", and "how" (definitions in Table ~\ref{tab:wh}). Each paper was coded based on these categories. We reviewed and refined the coding through multiple rounds to ensure accuracy and consistency. This process helped us create a well-structured dataset, which was used to summarize the trends, challenges, and areas of focus within digital labor research.

\subsection{Literature Comprehension}

This subsection provides an overview of the literature on digital work, synthesizing key findings from 143 reviewed papers. We examine \textbf{when} and \textbf{where} the research was conducted, \textbf{who} the primary stakeholders were, \textbf{what} challenges and dynamics were explored, \textbf{why} the work was undertaken, and \textbf{how} it was approached methodologically. While it is not possible to discuss every paper in detail, we present representative examples to illustrate prominent trends and insights. Our aim is to offer a clear and concise summary of the field, highlighting significant themes and areas that have received attention, as well as those that may require further exploration.

\subsubsection{When: 2015 to 2024}

Our selected set of literature focusing on labor and automation in AI through crowdwork spans a timeline from 2015 to 2024. Figure ~\ref{no_of_pap}(a) illustrates the number of papers identified yearly during our review. Our identified first paper was published in 2015, marking the early discussions on automation's implications for labor markets, particularly in the context of platforms like AMT and similar crowdsourcing systems. The most recent research is from 2024, reflecting ongoing and increasing interest in exploring how AI-driven automation affects labor practices, worker experiences, and policy responses worldwide.

\textbf{(a) 2015–2017: Early Explorations and Foundations.}
Between 2015 and 2017, 34 papers laid the foundation for understanding the impact of automation on digital labor. Research during this time primarily focused on platforms such as AMT, highlighting how human workers remained essential in supporting AI systems through tasks like data labeling, content moderation, and quality assurance \cite{2-irani2015difference, 3-salehi2015we, 4-irani2016stories, 5-hansson2016crowd, 11-mcinnis2016one, 12-kaun2020shadows, 20-chancellor2019human, 40-muldoon2023neither}. Early studies explored concepts like "ghost work" and raised concerns about the invisibility of workers behind seemingly automated processes. These years also marked the emergence of critical discussions on labor rights, ethical concerns, and the socio-economic challenges faced by workers, particularly in developed economies like the United States \cite{2-irani2015difference, 4-irani2016stories, 16-mcinnis2016running, 72-brawley2016work, 135-morris2015crowdsourcing}.

\textbf{(b) 2018–2020: Expansion of Global Perspectives.}
From 2018 to 2020, research expanded significantly, with 42 papers exploring the implications of AI and automation on workers in the Global South. This period marked an important shift in focus from primarily Western contexts to include regions like India, Southeast Asia, Sub-Saharan Africa, and Latin America. Studies highlighted the role of these regions in supporting global AI systems through digital freelancing and gig work, often under challenging conditions \cite{13-grohmann2021beyond, 21-wall2108artificial, 22-png2022tensions, 29-arora2016bottom, 30-van2020crowdsourcing, 65-altenried2020platform}. Researchers also began investigating the ethical implications of AI development and the role of policy in mitigating risks for vulnerable populations \cite{7-fox2020worker, 12-kaun2020shadows, 38-savage2020becoming, 86-anwar2020hidden, 137-bhatti2020general}.

The year 2019 was particularly prolific, with 23 papers published, the highest in any single year across our review. This surge can be attributed to the growing adoption of AI technologies in both developed and developing regions, alongside increased academic and industry interest in understanding their socio-economic impacts. Studies during this period often addressed how automation was reshaping traditional labor markets and creating new forms of precarious work \cite{19-inkpen2019human, 59-gandini2019labour, 60-howcroft2019typology, 83-goods2019your, 118-barbosa2019rehumanized}. The diversity of research topics in these years underscored the global scope of AI's impact and the complexity of its implications. 

\textbf{(c) 2021–2024: Focus on Policy, Ethics, and Regional Nuances.} The most recent years (2021–2024) saw a continuation of this momentum, with 67 papers delving deeper into the ethical, social, and economic dimensions of automation. Research in these years often focused on policy responses, regulatory frameworks, and the need for fairer working conditions on global platforms. Notable papers explored how automation affects marginalized communities, particularly in non-Western contexts, and emphasized the importance of culturally sensitive approaches to AI development and deployment \cite{10-irani2023algorithms, 23-okolo2023addressing, 26-chan2021limits, 42-do2024designing, 45-kim2024decoding, 66-wu2024gig, 67-sarker2024gender, 112-chen2024we}. The year 2021 saw significant contributions, with 15 papers exploring diverse topics such as labor platform regulation, worker conditions, and the intersection of automation with local economies. Similarly, in 2022 and 2023, research continued to highlight regional nuances, including case studies from Africa, India, and Southeast Asia, as well as discussions on the ethics of algorithmic systems in global markets \cite{24-naude2021crowdsourcing, 25-naude2022crowdsourcing, 37-hornuf2022hourly, 40-muldoon2023neither, 46-de2022understanding, 50-varanasi2022feeling, 88-hara2022understanding}. By 2024, research had evolved to address emerging challenges, including the need for more inclusive and equitable digital labor practices. Papers from this period increasingly emphasized the role of policy interventions and cross-regional collaborations in shaping the future of AI-driven automation \cite{42-do2024designing, 43-toxtli2024culturally, 45-kim2024decoding, 54-van2024understanding, 56-kincaid2024unconventional, 57-de2024we, 112-chen2024we, 139-bazaluk2024crowdsourcing}. These studies reflect the maturity of the field and the growing recognition of its global implications.


\begin{table}[t!]
\centering
\renewcommand{\arraystretch}{1.5} 
\setlength{\tabcolsep}{5pt}       
\begin{tabular}{|p{3.5cm}|p{11cm}|}
\hline
\textbf{Date filter} & 2015 $\leq$ year $\leq$ 2024 \\ \hline
\textbf{Search fields} & Everywhere (title, abstract, keyword, body) \\ \hline
\textbf{Quality Assessment} & Strongly reviewed research long and short research articles (manual filter) \\ \hline
\textbf{Search terms} & “Crowdwork,” “Ghost work,” “Crowdsourcing,” “Crowdsource,” “Crowdplatform,” and “Ghost Workers” \\ \hline
\textbf{Date of Query} & Initial: September 2024, First Update: November 2024, Second Update: January 2025 \\ \hline
\end{tabular}
\caption{Search variables that were used to find research articles on Google Scholar}
\label{search}
\end{table}

\begin{table}[t!]
\centering
\begin{tabular}{|p{1.2cm}|p{13.5cm}|}
\hline
\multicolumn{1}{|c|}{\textbf{Criteria}} & \multicolumn{1}{c|}{\textbf{Definition}} \\ \hline
\textbf{When} & 
``The publication year, ranging from 2015 to 2024.'' \\ \hline
\textbf{Where} & 
``The geographical distribution of studies, including regions like the USA, India, the Global South, and parts of Southern Africa.'' \\ \hline
\textbf{Who} & 
``The primary stakeholders discussed in the studies, including crowdworkers, platforms, clients, designers, and researchers.'' \\ \hline
\textbf{What} & 
``The primary focus of the research, such as labor dynamics and systemic challenges.'' \\ \hline
\textbf{Why} & 
``The ethical and societal impacts highlighted in the studies.'' \\ \hline
\textbf{How} & 
``The research methods used, such as qualitative, quantitative, or mixed-method approaches.'' \\ \hline
\end{tabular}
\caption{Criteria used to organize and analyze the literature on digital labor systematically. Definitions for "when," "where," "who," "what," "why," and "how" are provided, highlighting essential aspects like publication year, geographical focus, stakeholders, research themes, ethical considerations, and methodologies. These categories enabled us to identify key trends, challenges, and focal areas in the field.}
\label{tab:wh}
\end{table}

\subsubsection{Where: Geographical Distribution}

\begin{figure}[t!]
\centering
\includegraphics[width=0.99\textwidth]{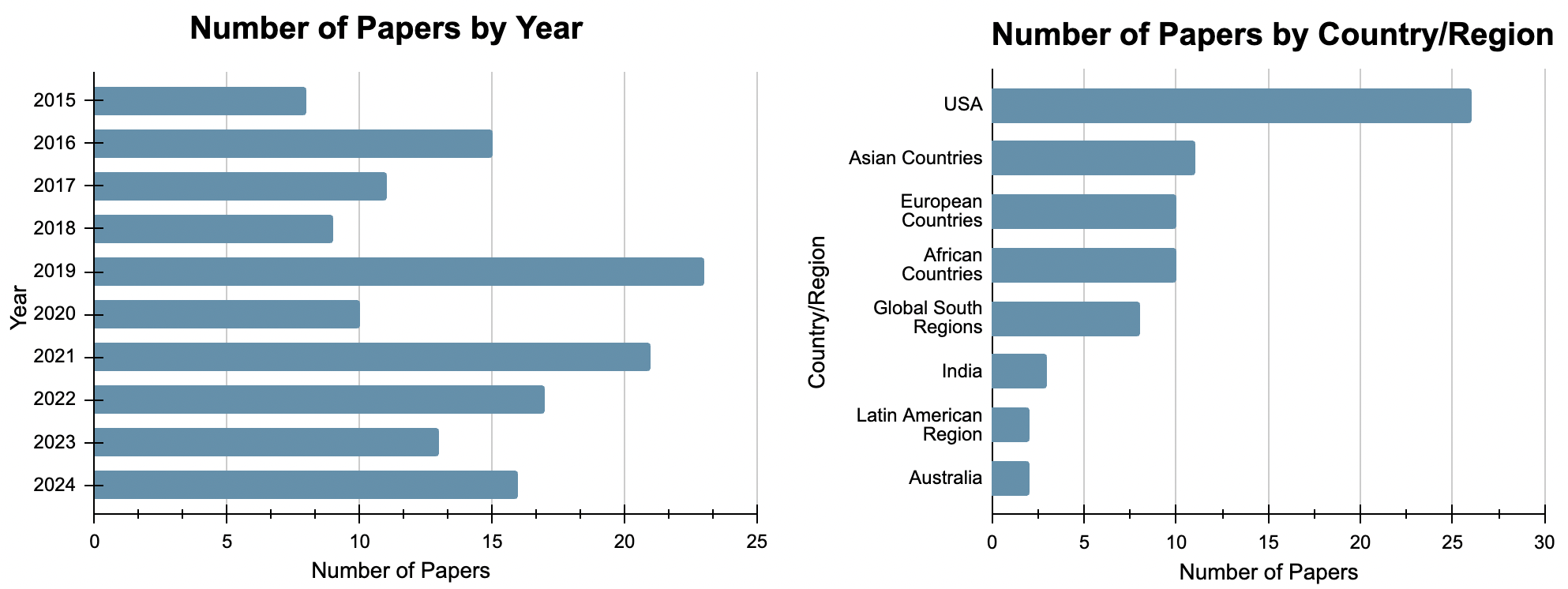} 
\caption{Yearly and Geographical Distribution of Reviewed Papers: (a) The yearly distribution of research papers published between 2015 and 2024. The y-axis represents the year, while the x-axis shows the number of papers published in each year. (b) The geographical distribution of the reviewed papers across regions. The y-axis represents the regions (e.g., USA, Asian Countries, etc.), and the x-axis shows the number of papers focused on each region.}
\label{no_of_pap} 
\end{figure}

Our selected papers covered diverse categories of geographical contexts, addressing labor and platforms in the United States, the Global South, and other regions. Figure~\ref{no_of_pap}(b) illustrates the geographical distribution of all our reviewed papers. A significant portion of the papers examine automation and digital labor networks in the United States, especially platforms based and widely used in the Unites States based platforms, such as AMT (e.g., \cite{1-silberman2018responsible, 14-willcocks2019hidden, 15-mcinnis2016taking, 32-toxtli2021quantifying, 33-whiting2019fair, 35-williams2019perpetual, 38-savage2020becoming, 39-rivera2021want, 41-abbas2022goal, 44-dutta2022mobilizing, 51-lascau2022crowdworkers, 53-hsieh2024supporting, 107-horton2016ghost}). A total of 26 papers specifically focus on contexts in the United States. The Global South receives significant attention, with eight papers focused on countries such as Brazil and India, African regions, Southeast Asia, and Sub-Saharan Africa. These studies explore labor practices, digital inequalities, and economic challenges faced by workers, among others \cite{13-grohmann2021beyond, 21-wall2108artificial, 22-png2022tensions, 23-okolo2023addressing, 26-chan2021limits, 29-arora2016bottom, 65-altenried2020platform, 75-wood2019good}. Specific mentions of Africa appear in six papers, with detailed investigations in regions such as South Africa, Kenya, Nigeria, Ghana, and Uganda \cite{24-naude2021crowdsourcing, 25-naude2022crowdsourcing, 30-van2020crowdsourcing, 79-anwar2021between, 84-wood2018workers, 86-anwar2020hidden}. Sub-Saharan Africa and Southeast Asia are highlighted in studies discussing digital labor conditions in these regions \cite{62-graham2017digital, 75-wood2019good}. India, as a critical hub for global AI labor, is a focal point in three papers \cite{10-irani2023algorithms, 29-arora2016bottom, 50-varanasi2022feeling}. Brazil appears in one paper, reflecting its emerging role in digital labor markets \cite{13-grohmann2021beyond}. Latin America, however, remains underrepresented, with only one paper specifically addressing this region \cite{30-van2020crowdsourcing}. Europe is represented in three papers, with studies focusing on labor markets and policy responses in countries like Germany and the European Union \cite{36-harmon2019rating, 87-tan2021ethical, 89-durward2020nature}. The United Kingdom is featured in four papers, often in the context of platforms like Prolific and minimum wage regulations \cite{40-muldoon2023neither, 41-abbas2022goal, 55-salminen2023fair}. Two papers specifically address China, focusing on platforms like Meituan and Didi \cite{69-xiang2024judging, 112-chen2024we}. Bangladesh appears in one paper, highlighting its role in digital freelancing \cite{67-sarker2024gender}. Australia is discussed in two papers, reflecting the growing gig economy in this region \cite{58-stewart2017regulating, 83-goods2019your}. Two papers also highlight Germany, focusing on Berlin’s urban gig economy and broader labor policies \cite{85-altenried2024mobile, 89-durward2020nature}. This geographical diversity underscores the global scope of automation and digital labor research while highlighting gaps, such as the limited attention given to Latin America and specific parts of Asia.

\subsubsection{Who: Stakeholders}
\begin{figure}[t!]
\centering
\includegraphics[width=0.75\textwidth]{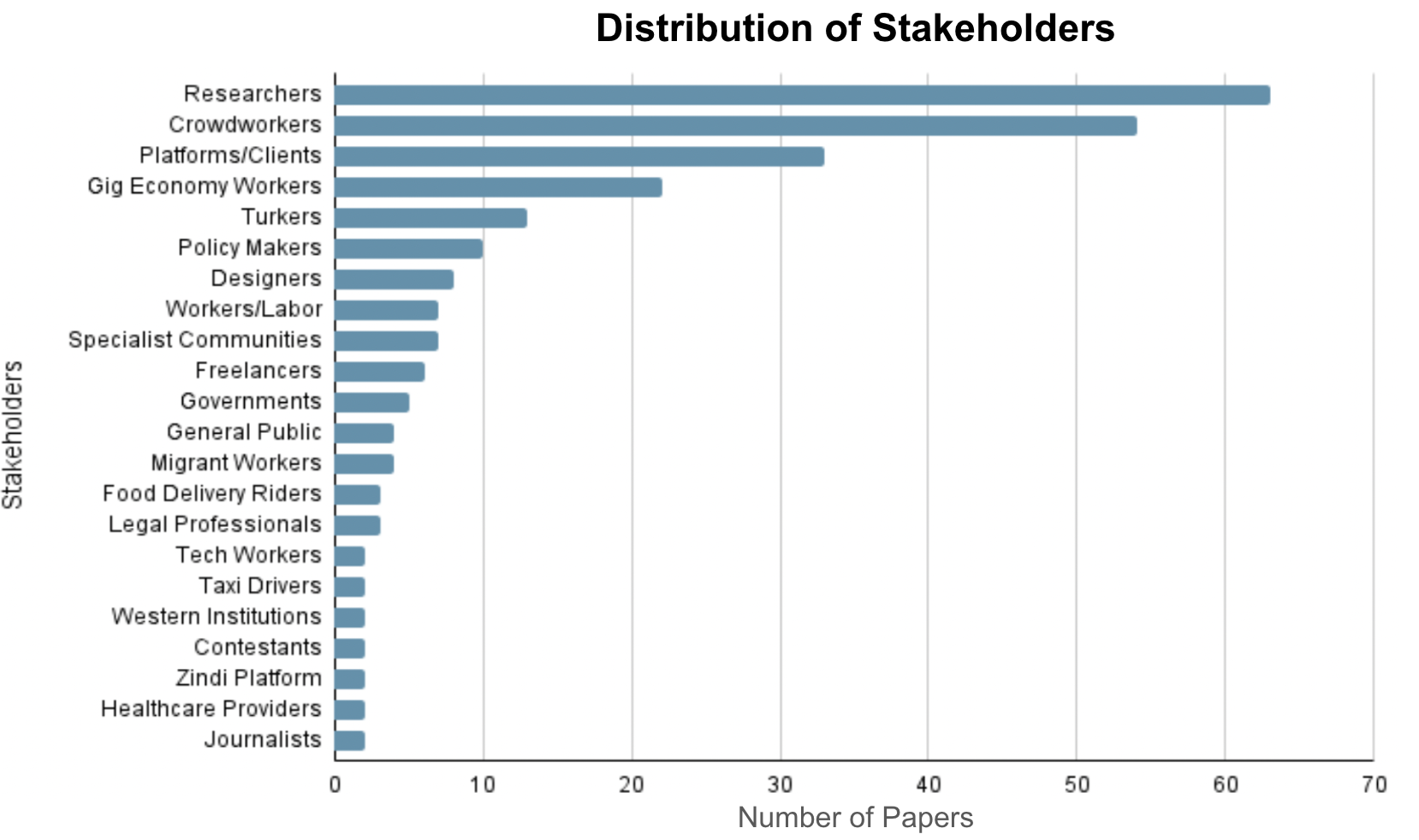} 
\caption{Breakdown of stakeholders in the crowd work and AI ecosystem, as identified in our reviewed papers. Stakeholders include crowdworkers, platforms and clients, designers, researchers, and other contributors, highlighting their roles and interactions }
\label{fig:papers_by_category} 
\end{figure}

The reviewed papers highlight various key players in the crowdwork and AI ecosystem. These include crowdworkers, platforms and clients, gig workers, researchers, and other stakeholders. Figure ~\ref{fig:papers_by_category} illustrates the distribution of stakeholders mentioned across all the papers reviewed in this study. Below is a detailed breakdown of each group's role and contributions based on the studies reviewed.

\textbf{Researchers: } Researchers investigating crowdwork, its ecosystem, and its implications are discussed in 63 papers (e.g., \cite{1-silberman2018responsible, 3-salehi2015we, 133-thuan2016factors}). These studies cover topics like worker conditions, platform design, wages, and ethical concerns. Some researchers also develop tools to support workers or analyze how platforms manage labor \cite{43-toxtli2024culturally, 44-dutta2022mobilizing, 73-gol2019crowdwork}.

\textbf{Crowdworkers: } A significant portion of the papers (54 papers) focus on crowdworkers, the individuals performing microtasks on platforms like AMT, Upwork, and Prolific (e.g., \cite{2-irani2015difference, 8-su2021critical, 18-feldman2021we, 27-du2024ethical, 28-ettlinger2016governance, 64-rani2021digital, 68-giang2024corporate, 70-tubaro2020trainer, 73-gol2019crowdwork}). Crowdworkers come from diverse locations, including both developed and developing countries \cite{26-chan2021limits, 34-berg2018digital, 36-harmon2019rating, 67-sarker2024gender}. They perform tasks such as data labeling, content moderation, and survey participation, which are essential for training AI systems and supporting online platforms. These tasks are often repetitive and require significant time and effort, yet the workers typically face low pay, unfair task rejection, and a lack of benefits \cite{3-salehi2015we, 16-mcinnis2016running, 27-du2024ethical}. For example, studies highlight the struggles of novice crowdworkers who earn less than experienced "super Turkers" \cite{38-savage2020becoming}. Particularly, Global South faces heightened challenges due to weak worker protections, intensifying job insecurity \cite{67-sarker2024gender, 79-anwar2021between}.

\textbf{Platforms and Clients: } The platforms that connect crowdworkers with tasks and the clients who post these tasks are another significant focus, discussed in 33 papers (e.g., \cite{27-du2024ethical, 28-ettlinger2016governance, 30-van2020crowdsourcing, 31-chuene2015adoption, 48-sannon2022privacy, 49-alvarez2023understanding, 52-hulikal2022collaboration, 63-van2017platform, 91-alkhatib2017examining}). Platforms like AMT, Upwork, and Appen act as intermediaries, setting the rules for task distribution, payment, and dispute resolution \cite{13-grohmann2021beyond, 46-de2022understanding}. These platforms are often criticized for prioritizing efficiency and profit over worker well-being \cite{28-ettlinger2016governance, 63-van2017platform}. Clients, including businesses and researchers, benefit from cost-effective labor but are often detached from the challenges faced by workers \cite{33-whiting2019fair, 48-sannon2022privacy}. Some studies highlight how platforms encourage task commodification, treating workers as interchangeable and obscuring their contributions \cite{45-kim2024decoding, 61-rosenblat2016algorithmic}.

\textbf{Gig workers: } Gig workers, including food delivery personnel, rideshare drivers, and other freelance laborers, are extensively discussed in 22 papers (e.g., \cite{7-fox2020worker, 42-do2024designing}). These workers often face challenges similar to those of crowdworkers, such as precarious employment conditions, low pay, and limited labor protections. Their work is integral to the gig economy, yet they remain subject to the pressures of algorithmic management and fluctuating demand.

Beyond crowdworkers, researchers, and platforms, several other important stakeholders emerge in the ecosystem.  Eight papers explore the role of designers in creating tools or improving platform fairness (e.g., \cite{18-feldman2021we, 4-irani2016stories}). Designers have developed systems like Turkopticon, which allow workers to share feedback about unfair task practices \cite{18-feldman2021we, 64-rani2021digital}. These tools aim to enhance transparency and support workers in advocating for better conditions, though systemic challenges remain. Governments and policymakers also play a crucial role, as highlighted in 10 papers (e.g., \cite{36-harmon2019rating, 37-hornuf2022hourly}). They are responsible for creating regulations that protect workers' rights, establish fair practices for platforms, and ensure that labor standards are upheld in the evolving digital economy.

However, the effectiveness of these regulations often varies, reflecting differences in national priorities and enforcement mechanisms. Freelancers, another significant group, are mentioned in six papers \cite{46-de2022understanding, 49-alvarez2023understanding, 52-hulikal2022collaboration, 63-van2017platform, 84-wood2018workers, 77-rahman2021invisible}. These individuals provide specialized services on platforms like Upwork, offering graphic design, programming, and content creation skills. Despite having flexibility, freelancers still grapple with uncertainties related to income, client relationships, and platform policies. Additionally, designers and engineers—those who create and manage the algorithms and systems governing digital platforms—emerge as key stakeholders. Several papers \cite{76-mohlmann2021algorithmic, 78-zhang2022algorithmic, 141-tavanapour2019towards} emphasize their influence on shaping worker experiences. For example, they design tools to allocate tasks, rate workers, and mediate disputes. These individuals hold significant responsibility for ensuring that platforms are transparent, equitable, and worker-friendly, though their priorities often align more closely with the interests of platforms and clients than with those of the workers themselves. This diverse range of stakeholders underscores the complexity of the crowdwork and gig economy ecosystem, where each group's roles, responsibilities, and impacts are deeply interconnected.

\subsubsection{What: Labor Dynamics}
The dynamics of digital labor reveal significant challenges and imbalances in how work is structured and managed in the AI ecosystem. Workers on digital platforms often face low pay, poor treatment, and limited recognition for their contributions. At the same time, companies benefit from invisible, low-cost labor while presenting themselves as technology-driven innovators. In this section, we discuss three key aspects of labor dynamics: the treatment of digital labor, the issue of power imbalances and invisible work, and the need for equitable governance in the AI ecosystem.

\textbf{(a) Treatment of Digital Labor.} Our reviewed papers derived insights into the role of low-cost labor and power imbalances in the automation of AI systems. Crowdsourcing platforms, like AMT, are used by companies to complete small tasks that AI cannot yet handle. However, workers on these platforms often face poor treatment, including low pay and lack of communication \cite{1-silberman2018responsible, 2-irani2015difference, 9-tandon2022barriers}. Our review also sheds light on the automation of labor management, where algorithms control worker assignments, monitor their performance, and even determine suspensions, leading to concerns about fairness and transparency \cite{2-irani2015difference, 65-altenried2020platform, 80-newlands2021algorithmic}. These platforms tend to conceal the labor behind AI, treating workers as invisible contributors rather than valuable participants \cite{2-irani2015difference, 12-kaun2020shadows, 22-png2022tensions, 87-tan2021ethical}. Additionally, we highlight the collective action efforts of crowd workers to address ethical issues in their work environments. Platforms like Dynamo help workers come together, discuss problems, and push for better treatment, such as creating ethical guidelines for research conducted on crowdsourcing platforms \cite{3-salehi2015we, 7-fox2020worker, 53-hsieh2024supporting}. Despite these efforts, challenges remain in organizing workers due to digital labor's decentralized and precarious nature \cite{3-salehi2015we, 66-wu2024gig}.

\textbf{(b) Power Imbalances and Invisible Work. }Power imbalances between workers and employers are a recurring theme in the reviewed literature. Workers often have little agency, with companies benefiting from decentralized, low-cost labor while presenting themselves as tech-driven rather than labor-driven enterprises \cite{2-irani2015difference, 4-irani2016stories, 5-hansson2016crowd}. Furthermore, design practices often portray designers as creative innovators while reducing crowd workers' roles to simple task completers, minimizing their contributions \cite{4-irani2016stories, 6-irani2018design, 35-williams2019perpetual}. We also found the hidden nature of labor in the digital economy. Workers performing essential tasks, such as data cleaning for AI training, often remain unseen \cite{12-kaun2020shadows, 13-grohmann2021beyond}. This invisibility reflects a broader issue of undervalued labor in AI development, where workers from low-resource settings contribute to global systems without receiving proper recognition \cite{12-kaun2020shadows, 22-png2022tensions, 26-chan2021limits, 86-anwar2020hidden}. Working conditions and worker well-being are also discussed. Many workers face challenges such as job insecurity, work-life conflicts, and emotional stress  \cite{32-toxtli2021quantifying, 34-berg2018digital, 39-rivera2021want}. Some rely on resilience strategies to navigate precarious environments, while others struggle with the physical and emotional toll of hidden labor \cite{46-de2022understanding, 75-wood2019good, 83-goods2019your}.

\textbf{(c) Equitable AI Governance. }Finally, we investigated the governance of AI in the Global South. There is a growing call for more equitable AI governance, where underrepresented regions are not merely included as labor sources but have a role in shaping AI technologies and policies. This includes addressing the risks of power concentration in Western countries and promoting more balanced participation from Global South actors in AI leadership \cite{21-wall2108artificial, 22-png2022tensions, 60-howcroft2019typology}. Transparency tools and participatory governance mechanisms are highlighted as potential solutions \cite{53-hsieh2024supporting, 113-hosseini2016crowdsourcing, 141-tavanapour2019towards}.

\subsubsection{Why: Ethical Impacts}

The ethical impact of AI automation and digital labor has been a central topic of research over the years. Early studies focused on understanding the challenges faced by workers, such as underpayment and lack of protections. Over time, researchers expanded their focus to explore solutions for improving fairness, well-being, and equity in digital work. Below we discuss three key areas of research: the evolving understanding of worker conditions, the role of algorithmic management, and the emergence of collective action and governance as solutions for ethical challenges.

\textbf{(a) Evolving Focus on Worker Conditions.} The discussion about AI automation and its impact on workers has evolved significantly over time. Initially, researchers focused on understanding the working conditions of crowdworkers and identifying the challenges they face. Early studies highlighted issues like underpayment, lack of social protections, and power imbalances between platforms and workers  \cite{1-silberman2018responsible, 2-irani2015difference, 22-png2022tensions, 23-okolo2023addressing, 26-chan2021limits}. As awareness grew, more attention was directed towards designing better systems and tools to empower workers, improve fairness, and enhance overall working conditions \cite{3-salehi2015we, 4-irani2016stories, 13-grohmann2021beyond}. Over time, researchers began to explore the \textbf{economic implications} of invisible labor and advocate for fair compensation for crowdworkers \cite{32-toxtli2021quantifying}. This marked an important shift toward making labor more visible and addressing issues of wage fairness and underpayment, which remain critical challenges today \cite{2-irani2015difference, 5-hansson2016crowd, 33-whiting2019fair, 37-hornuf2022hourly, 55-salminen2023fair}. Efforts to improve worker conditions also expanded to include understanding the \textbf{motivations of crowdworkers} and \textbf{their career aspirations} \cite{34-berg2018digital, 39-rivera2021want}. Researchers investigated how tools and platforms affect workers' productivity, well-being, and ability to achieve goals, proposing design changes to support better outcomes \cite{35-williams2019perpetual, 38-savage2020becoming, 41-abbas2022goal, 42-do2024designing}. Studies also began addressing broader issues like cultural diversity and inclusion on digital platforms, recognizing the importance of tailored interventions to meet diverse needs \cite{43-toxtli2024culturally, 44-dutta2022mobilizing}.

\textbf{(b) Impact of Algorithmic Management.} In parallel, researchers examined the role of algorithmic management and its impact on workers' autonomy and job satisfaction. They highlighted challenges such as opaque evaluations and algorithmic control, which often create insecurity and reduce workers' ability to succeed \cite{61-rosenblat2016algorithmic, 76-mohlmann2021algorithmic, 77-rahman2021invisible, 82-wood2021algorithmic, 78-zhang2022algorithmic}. Some studies proposed redesigning these systems to prioritize worker well-being and fairness \cite{78-zhang2022algorithmic}. As digital labor platforms expanded globally, there has remained a growing focus on their \textbf{effects in different regions}, especially the Global South. Researchers explored how crowdwork could empower marginalized groups and provide economic opportunities while addressing challenges like education gaps and infrastructural limitations \cite{62-graham2017digital, 24-naude2021crowdsourcing, 86-anwar2020hidden}. However, these studies also critiqued the exploitation of vulnerable workers, particularly in the Global South, and called for ethical practices to mitigate these issues \cite{65-altenried2020platform, 86-anwar2020hidden}.

\textbf{(c) Collective Action and Governance. } Over the years, collaborative solutions and collective actions emerged as key themes. Digital laborers increasingly organized to share information, demand better conditions, and push for policy changes \cite{53-hsieh2024supporting, 71-cini2023resisting, 84-wood2018workers}. Researchers supported these efforts by developing tools and systems that enable collective action and foster community among workers \cite{38-savage2020becoming, 53-hsieh2024supporting, 141-tavanapour2019towards}. More recent studies have expanded the conversation to include \textbf{governance and policy}. They emphasized the need for transparent platforms, fair labor practices, and worker protections through better governance mechanisms \cite{46-de2022understanding, 47-fan2023improving, 74-vallas2020platforms, 73-gol2019crowdwork}. Additionally, there has been a focus on ethical AI development and ensuring fairness in algorithms used by these platforms  \cite{19-inkpen2019human, 69-xiang2024judging, 118-barbosa2019rehumanized}. In summary, the timeline of research on AI automation and crowdwork reflects a growing understanding of the challenges workers face and the opportunities for improvement. From identifying fundamental issues like \textbf{underpayment} and \textbf{exploitation} to proposing innovative solutions for empowerment and fairness, this body of work continues to evolve, offering valuable insights for building a more equitable digital labor ecosystem.

\subsubsection{How: Research Methods}

The reviewed papers employ various research methods to explore various aspects of digital labor and platform ecosystems. These methods span qualitative, quantitative, and mixed approaches, providing diverse perspectives on the challenges, opportunities, and impacts of digital work. Below, we summarize the key research methods used in the studies, highlighting their unique contributions to the field.

\textbf{(a) Qualitative Approaches.} The papers use a variety of research methods, with most focusing on qualitative approaches (e.g., \cite{39-rivera2021want, 42-do2024designing}). These include interviews, focus groups, case studies, literature reviews, and participatory design sessions to explore the experiences of workers, platform designers, and other stakeholders. For example, interviews with crowdworkers highlight challenges like low pay and limited rights \cite{3-salehi2015we, 4-irani2016stories}. Ethnographic studies provide deeper insights into the lives of workers on platforms like AMT \cite{3-salehi2015we, 85-altenried2024mobile}. Participatory research, where researchers collaborate closely with communities, has also been used to understand worker needs and develop solutions (e.g., \cite{53-hsieh2024supporting, 78-zhang2022algorithmic}). Some papers use both qualitative and quantitative methods to provide a fuller picture (e.g., \cite{30-van2020crowdsourcing, 31-chuene2015adoption}). For instance, studies on AI adoption in developing regions combine surveys with interviews to understand the challenges and benefits of these technologies \cite{24-naude2021crowdsourcing, 25-naude2022crowdsourcing}. Similarly, mixed-methods research examines cultural influences and worker experiences by combining data analysis with field experiments \cite{43-toxtli2024culturally}.

\textbf{(b) Quantitative Approaches.} Quantitative methods are used in many papers to study trends and impacts of automation on the economy and society (e.g., \cite{40-muldoon2023neither, 41-abbas2022goal}). These methods include surveys, statistical modeling, controlled experiments, and algorithm evaluations. For example, field studies on AMT measure the time workers spend on invisible labor, revealing unfair compensation practices \cite{32-toxtli2021quantifying}. Other studies use experiments to test new ways to improve task allocation and reduce worker risks \cite{43-toxtli2024culturally, 118-barbosa2019rehumanized}. Some papers also focus on critical analysis to examine how automation systems affect existing social structures (e.g., \cite{63-van2017platform, 66-wu2024gig}). These include theoretical and historical reviews to understand the impact of platform labor on workers’ rights and propose improvements.

\textbf{(c) Longitudinal and Meta-Analytic Studies.} Longitudinal studies track workers over time to observe trends and changes in their work lives \cite{45-kim2024decoding, 122-schlagwein2019ethical}. Meta-analyses and systematic reviews combine data from multiple studies to identify key trends, gaps, and future directions (e.g., \cite{37-hornuf2022hourly, 142-ghezzi2018crowdsourcing}). For example, a systematic review of over 100 papers synthesizes recommendations for improving platform design \cite{141-tavanapour2019towards}. Framework development papers propose new ways to understand or improve systems, like allocating microtasks fairly or reducing biases in crowdsourcing platforms \cite{60-howcroft2019typology, 118-barbosa2019rehumanized, 143-simperl2015use}. Experimental studies test specific hypotheses, such as how transparent wage information affects worker choices \cite{88-hara2022understanding} or how new task designs impact pay equity \cite{43-toxtli2024culturally}. 
In summary, the papers collectively offer diverse and complementary perspectives on digital labor, using a wide range of research methods to explore and address the challenges workers face in the age of automation.

\subsection{Insights into what is missing}
Our literature survey reveals different ethical and accountability aspects of the domain of digital labor, such as unfair and unstructured payment, compensation policies, poor treatment, and limited recognition, and additional challenges with power practiced by different categories of stakeholders. We have mapped the timeline and geographic concentrations of the scholarship. However, several recent yet pressing concerns are yet to be integrated in digital labor for crowdwork literature. For example, the scholarship is currently missing a holistic understanding and strategies for labor policy in cross-border infrastructure, geography based-AI policies for labor, and remote work statutes imposed by different countries. Therefore, this gap in the knowledge impedes the design of fair, accountable, and sustainable digital labor infrastructure. Addressing these challenges will require future research and design strategies focused on fairness, accountability, and sustainability across platforms, policies, and stakeholder interactions.


\section{DISCUSSION}
This paper examines how digital labor scholarship in AI and human-centered computing engages with crowd platform markets, the people who work on them, and the ethical challenges they face. Drawing on an in-depth review of 143 research papers published between 2015 and 2024, we reflect on how digital workers are studied, what issues are emphasized, and which areas remain underexplored. In this section, we discuss the main gaps and overlooked patterns in the literature, followed by key implications for industry, policy, platform design, and academic research. We conclude by outlining the limitations of this study and describing our future research directions.

\subsection{Gaps and Trends in the Literature}
Our review of studies from 2015 to 2024 highlights several important patterns in how digital labor has been studied. \textbf{First, the growth of this scholarship has a dominant focus on improving laborers' performances, with noticeable negligence of workers' rights.} Much of the research focuses on describing the tasks being performed, such as data labeling and content moderation \cite{3-salehi2015we, 4-irani2016stories, 7-fox2020worker, 12-kaun2020shadows, 13-grohmann2021beyond}. However, there is less attention on their working conditions, long-term well-being, and the systemic issues that keep them invisible. We found that most studies focus on invisible digital labor in the Global South, where low wages and limited opportunities for workers make this kind of work to be exported to these regions \cite{13-grohmann2021beyond, 21-wall2108artificial, 22-png2022tensions, 23-okolo2023addressing, 26-chan2021limits}. Yet, these studies rarely address workers' lack of training and support, their consciousness of being discriminated against with wages and visibility, and how to ensure they are treated with fair wages and labor clauses so that they can contribute better to the global digital labor market. For example, while researchers often talk about improving wages, they do not explore practical policy design solutions for giving workers better recognition or representation in the AI industry.

\textbf{The second gap in the literature is the growing concern about the unintended and unpaid digital labor that platforms impose on users.} The writers of this manuscript investigated five dominant gig-labor platforms and found that multiple of them require users to watch advertisements or perform small tasks before they can access content or services, effectively turning these interactions into unpaid labor, which is commensurable to the findings of some rare existing research \cite{dis-hesmondhalgh2010user, dis-yazdanipoor2022digital}. It is unclear from the current trajectory of research and platform disclosures what the platforms do with the data associated with the unintended and unpaid jobs, and how that set of data is feed to algorithms. Another concern relates to the unequal burdens faced by users in the Global South when interacting with AI-driven systems \cite{dis2-krishna2024artificial, 23-okolo2023addressing}. Individuals traveling to  Global South often face extra barriers, such as completing multiple captchas to access services like Google or Facebook \cite{dis2-alashwali2020exploring}. These regional inequalities in how AI systems operate remain underexplored, despite being a clear example of how digital labor and systemic bias intersect. Addressing these overlooked issues is crucial to create a more equitable digital ecosystem.

\textbf{The third gap is the lack of voices from these invisible digital laborers themselves.} Many studies describe their experiences, but few include direct input from the workers \cite{3-salehi2015we, 4-irani2016stories, 10-irani2023algorithms, 7-fox2020worker, 11-mcinnis2016one}. This limits our understanding of their needs and challenges, which are crucial for the market's growth, so that the market's future can be reliable for the current laborers and new laborers can confidently pick this as a career choice. Our findings suggest that researchers need to focus more on these perspectives to create solutions that are meaningful and effective. Finally, our review shows that most work in this area does not discuss plurality in ethics as a concern in digital labor practices. Concept words like fairness, exploitation, and transparency are often mentioned but not explored enough to suggest actionable changes, particularly for complicated task contracts when they are cross-border.

\subsection{{Implications}}

Based on our review of the literature, we identify several important implications for improving digital labor markets and making them more fair, and sustainable. These implications are grouped by the areas where they may be most useful: industry, policy, platform design, and academic research. We hope this can help different groups think about how to support digital workers more effectively.

\subsubsection{Implications for Industry}

Our review of the literature highlights several ways the digital labor industry can better support workers by improving fairness, visibility, and sustainability across the AI market. \textbf{One important implication for industry is recognizing invisible labor and making worker contributions visible.} The reviewed literature shows that many digital workers feel invisible in the systems they support. Their labor often remains unrecognized even though it is essential for the functioning of AI systems. Companies and platforms can do more to acknowledge workers and make their contributions visible. For example, previous work has highlighted that the contributions of crowdsource and ghost workers are often invisible and undervalued in AI systems and platforms \cite{gray2019ghost, 107-horton2016ghost, 108-stilgoe2022ghost}. Recognizing labor boosts morale and helps create a sense of purpose and value among workers. It also helps bridge the disconnect between front-end AI development and back-end labor processes, reinforcing that human contributions are foundational to machine intelligence.

\textbf{A second implication involves designing fairer compensation systems that reflect the realities of digital labor.} Many workers in the Global South experience unstable income and limited labor protections. Businesses that rely on digital labor should work toward fairer pay structures that match the time and effort required for tasks \cite{109-haralabopoulos2019paid, 10-irani2023algorithms, 13-grohmann2021beyond}. This includes considering unpaid time, like waiting for jobs or learning new tools. Fair pay can help create a more reliable and motivated workforce, which benefits companies in the long run \cite{111-berg2015income, 112-chen2024we}. Studies also suggest that clearer wage standards or minimum rates may reduce income volatility and lead to more consistent worker engagement \cite{55-salminen2023fair}.

\textbf{A third implication for industry is addressing systemic inequalities to promote more inclusive participation.} Companies should also support workers from underrepresented regions by reducing systemic biases. This includes supporting local languages and offering tasks based on regional strengths \cite{126-hadavi2022duality, 127-deng2016duality}. This is important to reduce inequality in global labor participation. Companies may also benefit from creating programs that support long-term worker growth, such as offering certifications or badges for completed tasks, which could help workers build reputations and access more complex opportunities over time. This investment in worker development may contribute to improved task quality and worker retention, which are essential for the sustainability of digital labor platforms.

\subsubsection{Implications for Policy}

Our findings highlight the lack of basic protections for digital workers. Current systems often leave workers without safeguards or ways to report unfair treatment. Workers can be suspended without notice, underpaid, or denied access to recourse. \textbf{One key implication for policy is the need to guarantee basic labor protections for digital workers.} Governments and policymakers should consider regulations that offer basic protections—like minimum wage standards, fair dismissal policies, and grievance mechanisms \cite{120-grossman2018crowdsourcing, 121-zou2018proof, 16-mcinnis2016running}. These safeguards are especially crucial for digital workers who depend on crowd platforms as a primary source of income, yet often fall outside traditional employment protections.

\textbf{A second implication involves improving transparency and oversight of algorithmic management systems.} These systems, such as task distribution or performance evaluation, can introduce bias. Some workers receive fewer tasks because of where they live or how their profiles are ranked \cite{117-zhang2017consensus, 119-eickhoff2018cognitive}. Policymakers and independent bodies can push for transparency in how these systems work and require regular audits \cite{118-barbosa2019rehumanized}. Regular third-party audits and mandatory fairness reports could ensure that platforms remain accountable for how their systems affect different categories of workers. Transparency mandates could also be supported with incentives for platforms that meet fairness benchmarks.

\textbf{A third implication is the need for international cooperation to ensure fair treatment of cross-border digital labor.} In cross-border settings, international collaboration may be needed to create shared standards for fairness and accountability \cite{g8-berg2019digital}. This is particularly important given the growing reliance on labor from countries with weaker regulatory environments. Agreements between nations or international bodies could help define ethical baselines for platform labor practices, including dispute resolution and cross-border wage parity.

\textbf{Finally, policy reforms could explore new legal categories that reflect the realities of platform-based work.} There is also a need to expand labor protections beyond traditional employment definitions. Since many digital workers operate as freelancers or independent contractors, existing labor laws often do not apply to them. New types of protections tailored to platform-based work could ensure fairer treatment while respecting the unique structure of digital labor ecosystems. For instance, governments could consider hybrid legal categories that allow workers to retain some flexibility while still being guaranteed minimum protections. Discussions around "dependent contractors" or "platform workers" in some countries could serve as starting points for these reforms.

\subsubsection{Implications for Platform Design}

Designers and developers of platforms play a key role in shaping workers' experiences. \textbf{One important implication for platform design is giving workers more control over how they interact with tasks and clients.} The literature shows that many workers feel powerless because they cannot choose tasks freely, communicate with clients, or influence how they are rated \cite{13-grohmann2021beyond, 116-ford2015crowdsourcing}. Adding features that allow workers to rate tasks, send feedback, and communicate directly with clients could help increase their autonomy. Supporting worker autonomy can also reduce frustration and make platform use more sustainable, particularly for new workers who may find the systems difficult to navigate.

\textbf{Another implication is involving workers in the design process to ensure platforms reflect their real needs.} The studies also suggest that workers are rarely included in designing the systems they rely on. Platforms can improve fairness and usability by involving workers in decision-making \cite{128-stewart2019clinician, 129-preece2016crowdsourcing}. This might include surveys, user testing, or pilot programs where workers can give feedback on new features before they are launched. Incorporating such participatory approaches can lead to better tools and help workers feel heard and respected. In some cases, studies show that co-designed systems tend to better reflect the needs of diverse worker populations, including part-time, marginalized, or novice workers.

\textbf{A third implication for platform design is addressing regional diversity in infrastructure, language, and digital literacy.} Platforms that operate across countries must also consider regional differences. Workers in low-connectivity areas or with limited education may need different types of support. For example, offering tutorials in local languages or reducing technical barriers can help ensure broader access \cite{17-mcinnis2017crowdsourcing, 25-naude2022crowdsourcing}. Additionally, interface and task design should reflect the needs of users with different levels of digital literacy. Design simplicity and local relevance can help workers engage more effectively and reduce dropout rates.

\textbf{Finally, platform features should help build community, trust, and accountability among users.} Platforms should also support community and collective action. Tools like forums or task rating systems help workers share experiences and support each other \cite{26-chan2021limits}. These systems can improve transparency and help workers build stronger networks. Platforms can also support transparent wage information, clearer communication channels, and systems for resolving disputes may reduce uncertainty and improve trust between workers and clients. Together, these features promote accountability and give workers more confidence in the platform’s fairness. Over time, this kind of trust-building may lead to higher retention, better task performance, and reduced conflict between stakeholders.

\subsubsection{Implications for Future Research}

The research community has a role in improving digital labor practices. Many papers focus on worker challenges, but few include direct input from workers themselves. \textbf{One implication for future research is to involve workers more directly in the research process.} Future research should prioritize involving workers in interviews, co-design workshops, or community studies \cite{13-grohmann2021beyond}. Hearing from workers directly will lead to more grounded and relevant insights. These approaches can also help researchers better understand the emotional, social, and financial pressures faced by workers, which are often missed in survey-based or purely observational studies.

\textbf{A second implication is the need to explore the long-term development of workers and the support systems that can enable their growth.} Many workers want to grow their skills but lack access to training or education. Researchers can study how learning opportunities like tutorials, certifications, or upskilling programs can impact worker outcomes. These studies can inform platform designs, public policies, and educational initiatives that better serve digital laborers. In addition, comparative research across countries and regions could highlight how local infrastructures and educational access influence worker success and well-being.

\textbf{Finally, research should also examine how ethical platform design and worker-led tools can contribute to fairness in digital labor systems.} Both academia and industry can support research on fair algorithms and equitable platform design. Studies can explore how participatory design impacts worker experience or how algorithmic fairness practices affect job distribution and pay \cite{11-mcinnis2016one, 128-stewart2019clinician}. These efforts can lead to better-informed platform policies and help create a more ethical AI ecosystem. Researchers can also explore the effectiveness of worker-led tools and communities, such as forums or cooperative networks, in helping workers build power and advocate for better conditions over time. This may also include longitudinal research into how such communities evolve, the kinds of support they offer, and how they influence worker identity and solidarity. Future studies might also explore how ethical labor practices can be included in computer science, HCI, and design curricula, so that upcoming developers are encouraged to create platforms that treat workers fairly and support ethical working conditions.

\subsection{Limitations and Future Work}

While our study offers important insights into the conditions and ethical concerns surrounding digital labor, it has several limitations that must be acknowledged. First, our review is based entirely on published literature. Although this allowed us to examine a broad range of studies and synthesize insights across the field, we did not include direct input from workers, platform developers, or other stakeholders. This means our findings may not fully reflect the lived experiences, priorities, and challenges of those actively engaged in digital labor ecosystems. Including these voices is essential for developing a deeper understanding of the social, emotional, and structural dynamics that shape platform work. Second, our selection process relied solely on Google Scholar. While it is a widely used academic search engine, relying on a single source may limit the range of perspectives included in the review.  Another limitation of our work is that it does not present new empirical findings. We did not conduct interviews, observations, or participatory fieldwork to supplement the literature review. As a result, our findings are based solely on secondary interpretations of existing studies. While this approach allowed us to map out key patterns and gaps across the field, it may miss emerging concerns or nuanced insights that only surface through direct engagement with affected communities. 

These limitations open several promising directions for future work. One of our key goals moving forward is to produce design recommendations based on both existing literature and the input of stakeholders involved in digital labor systems. We plan to engage with a range of actors, including digital workers, trainers, platform designers, and developers. Through interviews, co-design workshops, and participatory research, we hope to refine and validate actionable strategies that address the real-world needs of those participating in or shaping AI-driven labor platforms. We are particularly interested in building a more participatory approach to knowledge production in this space. Much of the existing digital labor scholarship is still shaped by researcher-led interpretations, which can unintentionally reproduce hierarchies between those who study labor and those who perform it. Our future work aims to challenge this dynamic by foregrounding the voices of platform laborers in identifying problems, shaping research questions, and evaluating possible interventions.

\bibliographystyle{ACM-Reference-Format}
\bibliography{sample-base}

\end{document}